\documentclass[sigconf]{acmart}
\def\BibTeX{{\rm B\kern-.05em{\sc i\kern-.025em b}\kern-.08emT\kern-.1667em\lower.7ex\hbox{E}\kern-.125emX}}
\renewcommand\footnotetextcopyrightpermission[1]{} 
\usepackage{nicefrac}
\usepackage{siunitx}
\usepackage{array,framed}
\usepackage{booktabs}
\usepackage{
  color,
  float,
  epsfig,
  wrapfig,
  graphics,
  graphicx,
  subcaption
}
\usepackage{textcomp}
\usepackage{setspace}
\usepackage{tablefootnote}
\usepackage{bchart}

\usepackage{latexsym,fancyhdr}
\usepackage{enumerate}
\usepackage{algorithm2e}
\usepackage{algpseudocode}
\usepackage{graphics}
\usepackage{xparse} 
\usepackage{xspace}
\usepackage{multirow}
\usepackage{csvsimple}
\usepackage{balance}
\usepackage{booktabs}

\newcommand{\eg}{\emph{e.g.,}\xspace}

\usepackage{
  tikz,
  pgfplots,
  pgfplotstable
}

\usepackage{xcolor}
\definecolor{hous}{HTML}{b88b4d}
\definecolor{green}{HTML}{79c561}
\definecolor{farming}{HTML}{ded94c}
\definecolor{trans}{HTML}{b4b4a9}
\definecolor{services}{HTML}{ff362e}
\definecolor{other}{HTML}{dbd4d3}
\definecolor{industry}{HTML}{db79c0}
\definecolor{water}{HTML}{7982db}
\definecolor{techinfra}{HTML}{303355}

\usetikzlibrary{
  shapes.geometric,
  arrows,
  external,
  pgfplots.groupplots,
  matrix
}

\pgfplotsset{compat=1.9}

\usepackage{mathtools,}

\DeclareMathAlphabet{\mathcal}{OMS}{cmsy}{m}{n}

\DeclareGraphicsExtensions{%
    .png,.PNG,%
    .pdf,.PDF,%
    .jpg,.mps,.jpeg,.jbig2,.jb2,.JPG,.JPEG,.JBIG2,.JB2}

\usepackage{etoolbox}
\makeatletter
\def\@copyrightspace{\relax}
\makeatother
\setcopyright{none}

\begin{document}
\fancyhead{}
\def\thetitle{COVID-19 Screening Using Residual Attention Network an Artificial Intelligence Approach}
\title{\thetitle}

\author{Vishal Sharma}
\affiliation{\normalsize{Department of Computer Science}}
\affiliation{\normalsize{Utah State University}}
\affiliation{\normalsize{Logan, Utah}}
\email{vishal.sharma@usu.edu}

\author{Curtis Dyreson}
\affiliation{\normalsize{Department of Computer Science}}
\affiliation{\normalsize{Utah State University}}
\affiliation{\normalsize{Logan, Utah}}
\email{curtis.dyreson@usu.edu}

\begin{abstract}
Coronavirus Disease 2019 (COVID-19) is caused by severe acute respiratory syndrome coronavirus 2 virus (SARS-CoV-2). The virus transmits rapidly; it has a basic reproductive number ($R_0$) of $2.2-2.7$. In March 2020, the World Health Organization declared the COVID-19 outbreak a pandemic. COVID-19 is currently affecting more than $200$ countries with $6$M active cases. An effective testing strategy for COVID-19 is crucial to controlling the outbreak 
but the demand for testing surpasses the availability of test kits that use Reverse Transcription Polymerase Chain Reaction (RT-PCR). In this paper, we present a technique to screen for COVID-19 using artificial intelligence. Our technique takes only seconds to screen for the presence of the virus in a patient. We collected a dataset of chest X-ray images and trained several popular deep convolution neural network-based models (VGG, MobileNet, Xception, DenseNet, InceptionResNet) to classify the chest X-rays. Unsatisfied with these models, we then designed and built a Residual Attention Network that was able to screen COVID-19 with a testing accuracy of $98$\% and a validation accuracy of $100$\%. A feature maps visual of our model show areas in a chest X-ray which are important for classification. Our work can help to increase the adaptation of AI-assisted applications in clinical practice. The code and dataset used in this project are available at \textit{https://github.com/vishalshar/covid-19-screening-using-RAN-on-X-ray-images}.

\end{abstract}
\keywords{COVID-19 screening, Residual Attention Network, Deep Learning, Machine Learning, Chest X-Ray}

\maketitle

\vspace{-5pt}
\section{Introduction}
\label{sec:intro}

In February 2013, some people in Guangdong province in China became infected with a severe acute respiratory syndrome virus (SARS-CoV)~\cite{ref:covid_intro_2}. Eventually, SARS was detected in about $8000$ patients across $26$ countries, the World Health Organization (WHO) reported $774$ deaths due to SARS~\cite{ref:WHO_2004}. In September 2012, a similar incident happened with the Middle East respiratory syndrome virus (MERS-CoV). There were $2494$ confirmed cases of infection with $858$ deaths due to MERS-CoV~\cite{ref:WHO_2013}. 

Both SARS and MERS pale in significance to the latest CoV outbreak concerning human health.
In November 2019 pneumonia-like cases due to unknown causes started to appear in Wuhan, China killing hundreds of people in the initial weeks.  In early 2020, the International Committee on the Taxonomy of Viruses (ICTV) declared the virus as Coronavirus Disease 2019 (COVID-19) caused by the SARS-CoV-2 virus~\cite{ref:covid_intro_2}. The reproductive number ($R_0$) of COVID-19 is $2.2-2.7$~\cite{covid_rate} higher than SARS coronavirus due to a S protein in the RBD region of SARS-CoV-2~\cite{ref:covid_discovery_3}. The highly transmissible virus quickly spread globally. By August 10, 2020 COVID-19 had been detected in 213 countries with six and a half million active cases and seven hundred and thirty thousand deaths. 
Due to a lack of medical supplies and staff, COVID-19 has overpowered the medical system of over 200 countries, and was declared a pandemic by  World Health Organization (WHO) on March 11, 2020.

Diagnosing who has COVID-19 can help curb its spread by quarantining those infected. Currently, the most widely used technique for detecting COVID-19 is with viral nucleic acid detection using Reverse Transcription Polymerase Chain Reaction (RT-PCR), which works by detecting viral RNA from sputum or a
nasopharyngeal swab~\cite{ref:covid_diagnose_test}. Unfortunately, there is a shortage of RT-PCR test kits~\cite{ref:test_shortage}. RT-PCR tests are also relatively slow, a test takes, at best, about four hours to complete, even in a highly controlled environment. RT-PCR tests also have a high false positive rate (about 30\%)~\cite{ref:jcs_system_covid_19}. A common notable symptom of COVID-19 patients is difficulty breathing~\cite{ref:related_work_118}.

Recent advances in the field of computer vision suggest the possibility of a faster, more widely available alternative for detecting COVID-19. A CT scan of a patient's chest has shown higher accuracy and sensitivity than a RT-PCR test for COVID-19 detection~\cite{ref:jcs_7}. Another study~\cite{ref:jcs_8} further validates at least 20\%  higher detection sensitivity from CT scans versus RT-PCR tests. Identifying who has COVID-19 using an imaging technique is relatively non-intrusive, uses widely-available X-ray or CT scanners, and can be used at a very early stage to diagnose COVID-19~\cite{ref:deep_learning_attention}. 

Deep neural networks have been successful in processing medical images~\cite{ref:deepL_success} and image processing in general, for instance in object detection~\cite{ref:DL_objectD}, image segmentation~\cite{ref:DL_image_segmentation}, and image classification~\cite{ref:DL_imageC}.
Deep learning techniques such as, convolutional neural networks (CNN) are popular for processing medical images, \eg for classification~\cite{ref:CNN_medical_classification} and segmentation~\cite{ref:CNN_medical_segmentation}. Advances in CNN models over the past few years have led to robust implementations such as VGGNet~\cite{ref:VGG}, Inception~\cite{ref:inception}, DenseNet~\cite{ref:dense_net}, Xception~\cite{ref:xception}, and MobileNet~\cite{ref:mobile_nets}. 
Recently, Attention Mechanism~\cite{ref:attention} which generates attention-aware features, based on spatial features has become popular in the fields of computer vision and image processing. 

\begin{figure}
    \centering
	\includegraphics[width=0.48\textwidth]{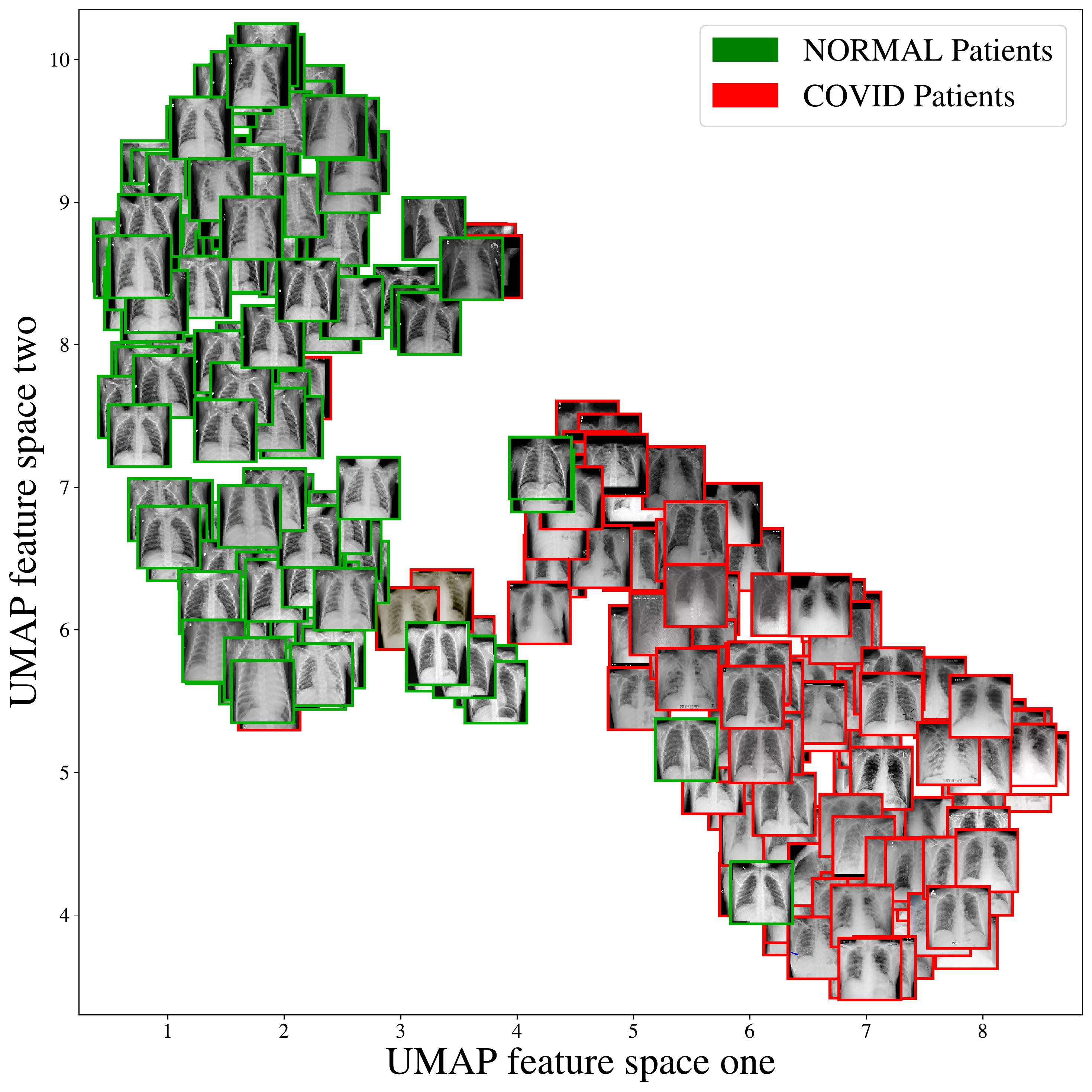}
	\caption{UMAP features of chest X-ray images of patients in our dataset}
	\label{fig:covid_umap}
\end{figure}

In this paper, we report on experiments with various deep learning models to detect and classify COVID-19 from chest X-ray images of patients. The models we use are VGG, ResNet, MobileNet, DenseNet, Xception, Attention, and Residual based CNN (Residual Attention Network). We first collected X-ray images of COVID-19 patients and people without the disease, which we call the ``normal" class. It is important to note that the normal class can have other illnesses, such as pneumonia. This image dataset is from a diverse population in terms of location, age, and gender.
To better understand the dataset, we analyzed it with a popular non-linear dimensionality reduction technique, Uniform Manifold Approximation and Projection (UMAP). 
The reduction produces a clear distinction between those with COVID-19 and those without (as shown in Figure~\ref{fig:covid_umap}).
Next, we split the dataset into training, testing, and validation sets using stratified folds.
We configured several deep learning models for optimal results and trained the models on the dataset. To improve the modeling, we designed and built our own Residual Attention Network with better feature extraction using custom designs of residual and attention block. Our model outperforms other models with 98\% accuracy on the test set and 100\% accuracy on the validation set. We extracted feature maps from our model and observed the Residual Attention Network detecting potential COVID-19 infected areas. The major challenge in this research is developing an effective model using only a very small dataset. We address this challenge by designing our Residual and Attention block to avoid overfitting. This paper makes the following contributions:
\begin{itemize}
\item a novel dataset of curated images for use in COVID-19 research,
\item a problem-specific, highly accurate classification model using Residual Connection and Attention Mechanism,
\item an explainable diagnosis using feature maps, and
\item a reproducible dataset and classifier; all of our code and data are in the public domain.\footnote{\url{github.com/vishalshar/covid-19-screening-using-RAN-on-X-ray-images}}
\end{itemize}
  
This paper is organized as follows. Section~\ref{sec:relwork} outlines related work and potential limitations. Section~\ref{sec:dataset} describes our dataset collection and analysis. Section~\ref{sec:approach} describes our approach and design of our Residual Attention Network. Section~\ref{sec:experiments} shows experimental configurations and reports results, and finally Section~\ref{sec:conclusion} presents conclusions and future work.

\section{Related Work}
\label{sec:relwork}

\begin{figure}
	\centering
	\resizebox{\linewidth}{!}{
		\begin{subfigure}[b]{0.5\textwidth}
			\begin{tikzpicture}
			\begin{axis}[
			xbar stacked, 
			xmin=0,
			ymin=1, 
			enlarge y limits=0.08,
			enlarge x limits=0.01,
			xmax=25,
			xmajorgrids=true,
			ymajorgrids=true,
			xlabel={\# of COVID-19 Patients},
			ylabel={Age Category},
			legend style={font=\fontsize{14.0}{8.5}\selectfont},
			legend pos=south east,
			ytick={1,2,3,4,5,6,7,8},
			ticklabel style = {font=\Large}, 
			label style={font=\Large}, 
			yticklabels={10-20, 20-30, 30-40, 40-50, 50-60, 60-70, 70-80, 80+},
			nodes near coords,
			every node near coord/.append style={font=\Large},
			scaled x ticks = false, 
			bar width=9pt, 
			grid style=dotted,
			width=.95\textwidth
			]
			\addplot coordinates {
				
				(1, 1) 
				(2, 2) 
				(6, 3) 
				(12, 4)
				(11, 5) 
				(14, 6) 
				(10, 7) 
				(1, 8)};	
			\addplot coordinates {
				
				(0, 1) 
				(2, 2) 
				(2, 3) 
				(4, 4)
				(11, 5) 
				(11, 6) 
				(5, 7) 
				(0,  8)  };	
			\legend{\strut Male, \strut Female}
			\end{axis}
			\end{tikzpicture} 
			\caption{\huge Distribution w.r.t Age }
			\label{fig:age_distribution}
		\end{subfigure}%
		\hfill
		
		\begin{subfigure}[b]{0.5\textwidth}
			\begin{tikzpicture}
			\begin{axis}[
			xbar, 
			xmin=0,
			ymin=0, 
			enlarge y limits=0.08,
			enlarge x limits=0.01,
			xmax=40,
			xmajorgrids=true,
			ymajorgrids=true,
			xlabel={\# of COVID-19 Patients},
			ytick={0,1,2,3,4,5,6,7,8,9,10,11,12,13},
			yticklabels = {Israel,Sweden,Iran,UK,Canada,Australia,USA,Vietnam,Taiwan,China,Spain,Italy},
			nodes near coords,
			ticklabel style = {font=\Large}, 
			label style={font=\Large}, 
			every node near coord/.append style={font=\Large},
			scaled x ticks = false, 
			bar width=7pt, 
			grid style=dotted,
			width=.95\textwidth
			]
			\addplot coordinates {
				(1, 0) 
				(1, 1) 
				(1, 2)
				(1, 3) 
				(1, 4) 
				(2, 5) 
				(4, 6) 
				(6, 7) 
				(12,8) 
				(13,9) 
				(17,10) 
				(34,11) };	
			\end{axis}
			\end{tikzpicture}
			\caption{\huge Distribution w.r.t Location}
			\label{fig:dataset_distribution}
		\end{subfigure}%
		\hfill
	}
	\caption{Distribution of our dataset with positive COVID-19 based on location, gender and age}
	\label{fig:dataset}
\end{figure}
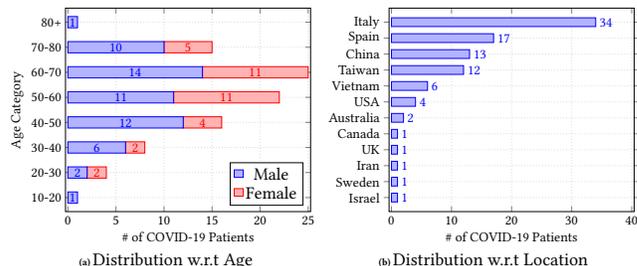

There are over 24,000 research papers on COVID-19 from well known sources like $bioRxiv$, $arXiv$ and $medRxiv$. More than 1,500 of these papers are peer reviewed~\cite{ref:related_number}. There are two recent review papers about using AI techniques in COVID-19 detection ~\cite{ref:background_1, ref:background_3}. 
We focus on research that uses deep learning for COVID-19 detection. 
Wang et al.~\cite{ref:related_work_103} used 1,065 chest CT scan images of COVID-19 patients to build a classifier using InceptionNet. They report an accuracy of $89.5$\%, a specificity of $0.88$, and a sensitivity of $0.87$. Xu et al.~\cite{ref:related_work_104} used 3D Convolution Neural Networks (CNNs) and reported an accuracy of $86.7$\%. Chen et al.~\cite{ref:related_work_105} segment the infected areas in CT scans using UNet++~\cite{ref:segmentation_unet++}. Using transfer learning and predefined models to classify COVID-19 in CT scans has also been researched, for instance using DenseNet \eg~\cite{ref:related_work_108, ref:related_work_128},  ResNet \eg~\cite{ref:related_work_121, ref:related_work_112}, and CNN \eg~\cite{ref:related_work_121, ref:related_work_108, ref:related_work_122}. Traditional machine learning (ML) methods of feature extraction and conventional ML algorithms for classification have also been used. Mucahid et al.~\cite{ref:related_work_120} used feature extraction techniques GLCM (grey level co-occurrence matrices), LDP (local directional pattern), GLRLM (grey-level run length matrix), and DWT (discrete wavelet transform), and using extracted features in a Support Vector Machine (SVM) for classification.  They report an accuracy of $99.68$\% in the best configuration settings. Alqudah et al.~\cite{ref:related_work_125} applied various ML techniques, such as SVM and Random Forest, and reported an accuracy of $95$\%. 

To the best of our knowledge, we are the first to design and build a Residual Attention Mechanism to extract spatial-aware features and perform the classification of a COVID-19 dataset. We obtained higher test and validation accuracy, precision, recall, sensitivity, and specificity than previous work. Additionally, most of the previous research used more sophisticated CT scan images which usually take $20$ to $30$ minutes to perform, but we use X-ray images that are faster to extract, about 10 minutes in most cases. X-ray machines are more widely available than CT scanners, and there are portable X-ray units that can be deployed anywhere, not just in medical facilities.

\section{dataset}
\label{sec:dataset}



We collected images only from public sources, which provide the data while maintaining patients' privacy.
The dataset of COVID-19 X-ray images comes from radiopaedia.org\footnote{radiopaedia.org}, the website of the Italian Society of Medical and Intervention Radiology\footnote{https://www.sirm.org/category/senza-categoria/covid-19/}. Cohen et al.~\cite{ref:covid_dataset} scraped the images from the website using PDF processing tools. We selected 120 images of patients with COVID-19, specifically, we selected all \textit{Posterioranterior} (PA) images.  The PA view has an anterior aspect in which the ribs are much clearer than the \textit{Anteriorposterior} or \textit{Lateral} view X-ray. To collect images of non-COVID-19 (which we call normal) chests we randomly selected images from a repository collected by Mooney et al.~\cite{ref:kaggle_normal}. In this repository, there are chest X-ray images of patients with pneumonia as well as of healthy patients. We extracted $119$ PA view X-ray images of normal patients. In total, our dataset has $239$ images with $120$ from COVID-19 patients and $119$ from normal patients.

\subsection{Dataset Statistics}
A wide distribution in a dataset is an important factor for training a deep learning model, since training on narrowly distributed data may lead to a biased model due to a failure to generalize the classification features.  Figure~\ref{fig:dataset} depicts our dataset distribution, with respect to location, gender, and age of patients. The figure shows that our image collection comes from $62$ male and $39$ female patients with a normal age distribution that is shifted towards elderly patients. These patients are from twelve countries, however, patient ethnicity was not made public.


\textbf{Age:} The dataset has patients from age $12$ to $84$, with an average age of $57.33$ years. Figure~\ref{fig:age_distribution} shows the amount of samples in the age category as well as the gender count in each category. Our dataset has a wide range of patients in terms of their ages.

\textbf{Location:} Figure~\ref{fig:dataset_distribution} shows the location of patients. The location is an important attribute, since a model trained on data from only one country may become biased. Greater variation in the training data can help generalize a deep learning model. Our dataset has images of patients from twelve countries. 



\subsection{UMAP Exploration} We applied the Uniform Manifold Approximation and Projection (UMAP) technique, which is a non-linear dimensionality reduction technique, to the images. The feature space of UMAP is found by searching for a low-dimensional projection of data which is the closest equivalent to the real data using a fuzzy topological structure. We prepared the dataset for UMAP by performing a standard image pre-processing, as described further in Section~\ref{sec:data_preprocessing}. 
Figure~\ref{fig:covid_umap} shows the result of the UMAP. In the figure, the X-ray of a normal patient has a green bounding box while that of a COVID-19 patient has a red bounding box.
The figure shows two clusters, one dominated by COVID-19 images, the other by normal images.

\section{Approach}
\label{sec:approach}
In this section, we explain our custom Residual block, Attention block, and how we used them to design and build a Residual Attention Network, which extends the original Residual Attention Network~\cite{ref:residual_attention}. 

\subsection{Residual Block}

\begin{figure}
	\centering
	\includegraphics[width=0.42\textwidth]{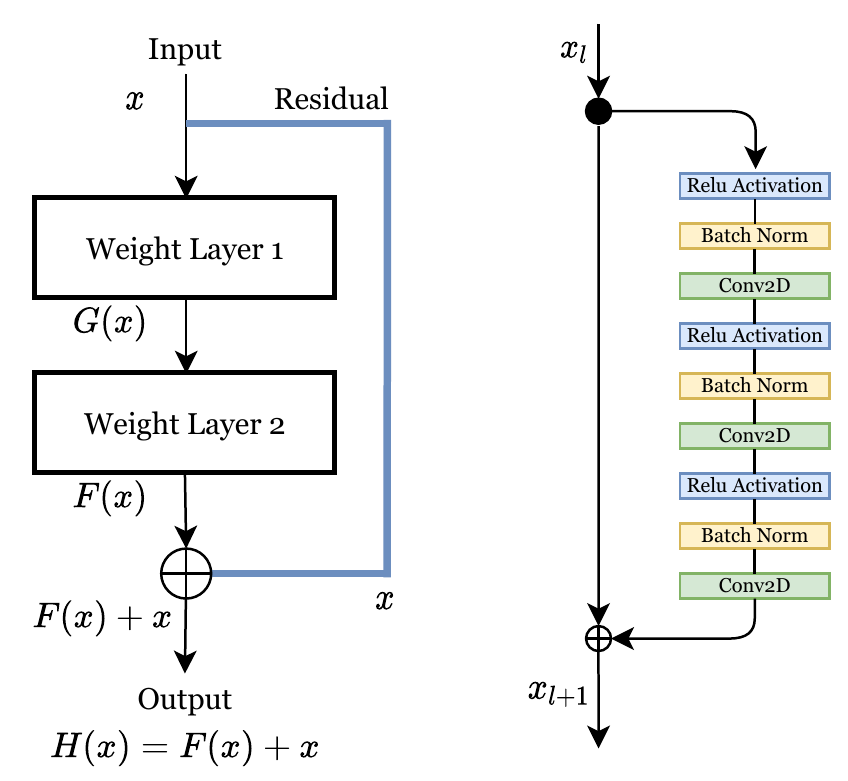}
	\caption{Image on the left shows a Residual block and on the right shows full pre-activation used as our Residual block}
	\label{fig:residual_block_dia}
\end{figure}

Deep convolution networks have revolutionized the field of image classification. Advancements in algorithms and hardware networks have increased the ability to add layers to a deep convolution network. With the increase in the depth of the network, it becomes harder to train a neural network because of vanishing gradients. Networks with too many layers become highly unstable as the value of gradients approaches zero in early layers. Every additional layer gradient value becomes smaller and eventually insignificant. Vanishing gradients degrade the performance of the network and adding more layers only exacerbates the problem. To solve the vanishing gradient problem, Kaiming He et al.~\cite{ref:res_net} proposed \textit{residual connections}. A residual connection merges the output of a layer with the input of a previous layer, which ensures that gradient values do not suddenly vanish. 
As shown in Figure~\ref{fig:residual_block_dia}, on the left is a residual block with a residual connection. 

A deep learning model in general tries to learn a mapping function $H(x)$ from an input $x$ to output $y$,
\begin{equation}
	H(x) = y
\end{equation}
In a residual block, instead of learning a direct mapping, it uses the difference between the mapping of $x$ and the original input $x$, 
\begin{equation}
 F(x) =  H(x) - x
\end{equation}
re-arranging gives, 
\begin{equation}
H(x) = F(x) + x 
\end{equation}
residual block learns the residual $F(x)$ with given $x$ as an input and $H(x)$ as the true output. This technique helps when increasing the depth of a neural network. 

Our experiments with arranging residual block for optimal gradient flow showed full pre-activation with batch normalization gives the best results, which was also suggested by Kaiming He et al.~\cite{ref:res_net}. Our full pre-activation block is shown in the right of Figure~\ref{fig:residual_block_dia}, where we use Relu activation, batch normalization and a 2D convolution layer stacked three times. The sequence and stacking of our block are  different than originally proposed~\cite{ref:residual_attention}, which was batch normalization, Relu activation and convolution layer stacked twice. From our experiments, we observed that batch normalization after Relu activation performs better. The reason for better performance happened when input features for a layer are negative, in the network they would have been truncated using non-linearity activation function, Relu, before batch normalization. Performing batch normalization prior to Relu activation will include these negative values in feature space.
\subsection{Attention Block}
The attention mechanism has become a very popular technique in natural language processing, image processing, and computer vision. The attention mechanism can generate attention-aware features and features that can be extracted based on spatial, context, or channel aware-features. It also learns the importance and correlations among features. Using visual attention in an image classification task helps determine important image regions and their correlations. The presence or absence of image regions is critical to classification. In our case, these regions are evidence of COVID-19 infection in chest X-ray images.
Our attention module consists of two branches: a ``trunk'' branch $T(x)$ with two stacked residual blocks and an encoder-decoder ``mask'' branch $M(x)$. 

\begin{figure}
	\centering
	\includegraphics[width=0.49\textwidth]{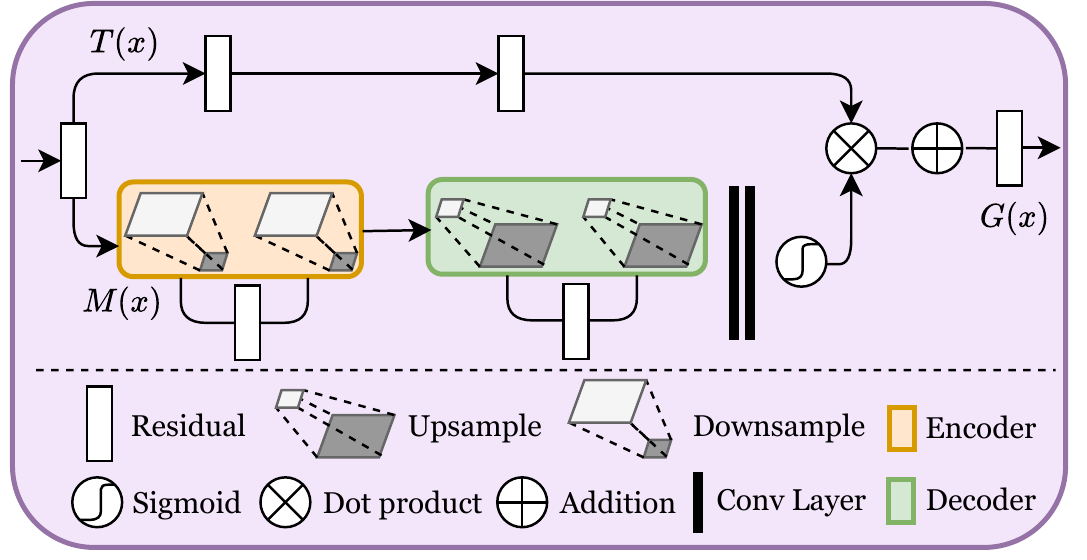}
	\caption{Design of Attention Block}
	\label{fig:attention_block_dia}
\end{figure}

The encoder in the mask branch consists of downsampling using max-pooling followed by residual connection and downsampling again. The encoder acts as an input reducer. The decoder consists of upsampling using bilinear interpolation. In the original Residual Attention Network~\cite{ref:residual_attention} there is only one upsample but, in a literature survey we found that the performance of a Residual Attention Network can be increased by increasing the number of up-sampling layers. We extended the Attention block by adding two upsample layers in our model. The encoder and decoder are followed by two convolution layers and a sigmoid activation as displayed in Figure~\ref{fig:attention_block_dia}. The trunk branch consists of two stacked residual blocks that perform feature processing.

The final output of the module is
\begin{equation}
G(x) =  (1 + M(x)) \ast T(x)
\end{equation}
adding 1 to the equation ensures that in case of mask branch with zero output the trunk branch computation passes through, which dampens the susceptibility to noisy data.

\subsection{Residual Attention Network}

Our Residual Attention Network is built by multiple stacks of our basic unit module. The stacking of blocks is designed for optimal performance and to prevent overfitting. The Attention block is designed to explore fine-grained feature maps since COVID-19 infections could be a fine detail in an X-ray. There are two major attention categories: \textit{Soft} and \textit{Hard}, we use \textit{Soft} attention to learn alignment for several patches. 

The Residual block captures high-level features and provides input to Attention block. The Attention module generates specialized low-level features on Residual input. It divides an image into a few high-level features and from those features extracts several low-level features. We stack Residual and Attention layers alternatively three times. The Residual layer extracts high-level features from the input image which are then passed to the Attention block, which extracts low-level features. These low-level features become an input to the next Residual block. It works as both a top-down and bottom-up approach, the top-down network produces dense features and the bottom-up one produces low-resolution feature maps. Our architecture is shown in Table~\ref{tab:residual_architecture}. 
This technique has proved successful in image segmentation. Our scenario is very similar to segmentation where we try to identify low-level patches of COVID-19 infections in a chest X-ray. 

\begin{table}[h]
	\centering
	\begin{tabular}[t]{>{\normalsize}l>{\normalsize}c>{\normalsize}c>{\normalsize}c}
		\hline
		\textbf{Layers}&\textbf{Output Size}&\textbf{Kernel Size}\\
		\hline
		
		Convolution 2D 	& 112$\times$112  & (5$\times$5), $p$=same 		\\
		MaxPool 2D 		& 56$\times$56	  & (2$\times$2), 2 		\\

		Residual Block 	& 56$\times$56 	  & $\begin{pmatrix}
											1\times1 & 32\\
											3\times3 & 32\\
											1\times1 & 128
											\end{pmatrix}$ 		 \\			
		Attention Block & 56$\times$56    & Attention$\times$1 \\

		Residual Block 	& 28$\times$28    & $\begin{pmatrix}
											1\times1 & 128\\
											3\times3 & 128\\
											1\times1 & 256
											\end{pmatrix}$ 		 \\		
		Attention Block & 28$\times$28    & Attention$\times$1 \\

		Residual Block 	& 14$\times$14    & $\begin{pmatrix}
											1\times1 & 256\\
											3\times3 & 256\\
											1\times1 & 512
											\end{pmatrix}$ 		 \\		
		Attention Block & 14$\times$14    & Attention$\times$1 \\

		Residual Block 	& 7$\times$7 	  & $\begin{pmatrix}
											1\times1 & 512\\
											3\times3 & 512\\
											1\times1 & 1024
											\end{pmatrix}$ 		 \\			
		Residual Block 	& 7$\times$7 	  & $\begin{pmatrix}
											1\times1 & 1024\\
											3\times3 & 1024\\
											1\times1 & 1024
											\end{pmatrix}$ 		 \\										 
		Residual Block 	& 7$\times$7 	  & $\begin{pmatrix}
											1\times1 & 1024\\
											3\times3 & 1024\\
											1\times1 & 1024
											\end{pmatrix}$ 		 \\

		AvgPooling 2D 	& 1$\times$1 	  & (7$\times$7) 		 \\	
		FC, Softmax 	& \multicolumn{2}{c}{2} 				 \\

		\hline
		Depth 			& \multicolumn{2}{c}{115} 				 \\
		\hline
	\end{tabular}
	\vspace{15pt}
	\caption{Residual Attention Network architecture}
	\label{tab:residual_architecture}
\end{table}

\section{experiments}
\label{sec:experiments}

This section presents an experimental evaluation of our model. We start with data preprocessing and stratified data split for training, testing, and validation. We describe the experimental setup, our models, and their configurations, and we show the results. 

\subsection{Data Preprocessing}
\label{sec:data_preprocessing}

Data preprocessing plays an important role in training a deep learning model. Previous research has highlighted the impact of preprocessing on model performance. 
Images in our dataset have different sizes, so we first start with standardizing the image size to $224\times224$ pixels. The collected images also have different color patterns, so we normalize the color pattern to RGB. Lastly, we normalize the maximum intensity of a pixel to $255$ (lowest is $0$). Our dataset has two different classes, labeled \textit{COVID}, and \textit{Normal}; during training, we use a label binarizer to convert them to one-hot encoding.

\subsubsection{Data Split}
The dataset is split into training ($70$\%), testing ($20$\%), and validation ($10$\%) sets. We use stratified splits which ensures that each split has an equal number of samples from each class as shown in Table~\ref{tab:dataset_split}. We also add a random rotation of $15^{\circ}$ to images, which adds more stability to our model during training. 

\begin{table}[ht]
	\centering
	\begin{tabular}[t]{>{\normalsize}l>{\normalsize}r>{\normalsize}r>{\normalsize}r}
		\textbf{}&\textbf{COVID}&\textbf{Normal}&\textbf{Total} \\
		\hline
		Training 	 & 84  & 83 & 167 \\
		Testing 	 & 25  & 25 & 50 \\
		Validation   & 11  & 11 & 22 \\
		\cline{2-4} 
		& \small{120} & \small{119} & \small{239} \\
		\hline
	\end{tabular}
	\vspace{15pt}
	\caption{COVID dataset splits using stratified sampling}
	\label{tab:dataset_split}
\end{table}

\subsection{Experimental setup}
All experiments were carried out on a computer with Intel i$7$ $5820$k, Nvidia $1080$ti, $16$GB of RAM running Ubuntu $18.04$ OS. We use Python $3.7$ and its libraries (sklearn, tensorflow, keras) to write and train the deep learning models. All of the networks were trained on Nvidia $1080$ti ($3584$ CUDA cores and $11$GB DDR$5$ RAM) with CUDA and cuDNN configured for performance enhancement. 

\subsubsection{Baseline} 
We use the following models as baseline for our Residual Attention Network: 

\textbf{VGGNet}: VGGNets were introduced in 2014 with the intention to perform state of the art image classification with the least depth of CNN possible. We use both VGG16 and VGG19 in our experiments.

\textbf{ResNet}:  ResNet introduced the concept of residual connection to solve the vanishing gradient problem. Deciding a kernel size for ResNet is hard and that's why we use InceptionResNet a variant of ResNet, which uses multiple size kernels within the same layer. 

\textbf{Xception}: Xception introduced the concept of depthwise separable convolution to reduce the number of the parameter without loss of performance. 

\textbf{MobileNet}: In addition to depthwise convolution, MobileNet introduced pointwise convolution to reduce the number of parameters in order of $100$ to $1000$. We specifically use MobileNet and its variant MobileNetV2.

\textbf{DenseNet}: DenseNet extends the idea of residual from ResNet, but instead of learning residual (the difference between previous and current layer) it proposes to merge the output of the previous and current layer. We use DenseNet121 and DenseNet201 in our experiments. 

\textbf{NASNet}: NASNet is Neural Architecture Search Network. They are a family of models designed to learn model architecture automatically on the dataset of interest. In our case \textit{imagenet} was used as dataset to design NASNet.

\textbf{Vanilla Residual Attention Network (RAN)}: To compare our work with existing state-of-the-art Residual Attention Network, we also implemented Attention-56 which was proposed in~\cite{ref:residual_attention}. 

\subsubsection{Evaluation Metric}
To evaluate the performance of our models, we use the most commonly used performance metrics for deep learning, namely, sensitivity, specificity, precision, recall, and accuracy. Their value range is $[0,1]$: higher is better. The metrics are given below, where TP is true positive, FP is false positive, TN is true negative, and FN is false negative.

\vspace{5pt}
{
\centering
Sensitivity (Sens) $= \frac{\text{TP}}{\text{TP + FN}}$ 
\\
\vspace{5pt}

Specificity (Spec) $= \frac{\text{TP}}{\text{FP + TN}}$ 
\\
\vspace{5pt}

Precision (Prec) $= \frac{\text{TP}}{\text{TP + FP}}$ 
\\
\vspace{5pt}

Recall (Rec) $= \frac{\text{TP}}{\text{TP + FN}}$ 
\\
\vspace{5pt}

Accuracy (Acc)$= \frac{\text{TP+TN}}{\text{TP+FP+TN+FN}}$ 
\\
\vspace{5pt}
}

\begin{figure*}[t]
	\centering
	\begin{minipage}{0.49\linewidth}
		\centering
		\includegraphics[width=\linewidth]{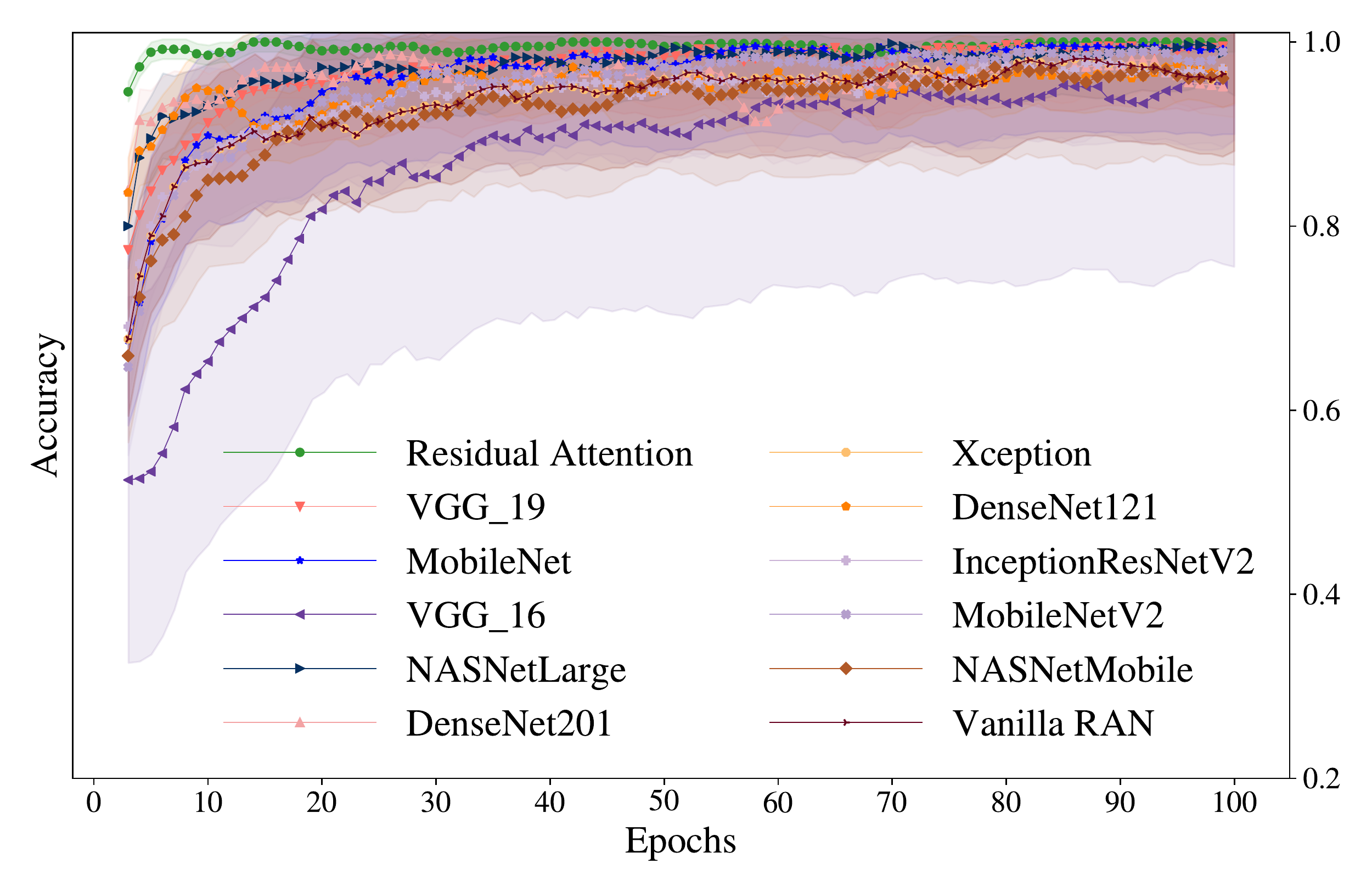}
		Accuracy over epochs of models on training set
	\end{minipage}
	\begin{minipage}{0.49\linewidth}
		\centering
		\includegraphics[width=\linewidth]{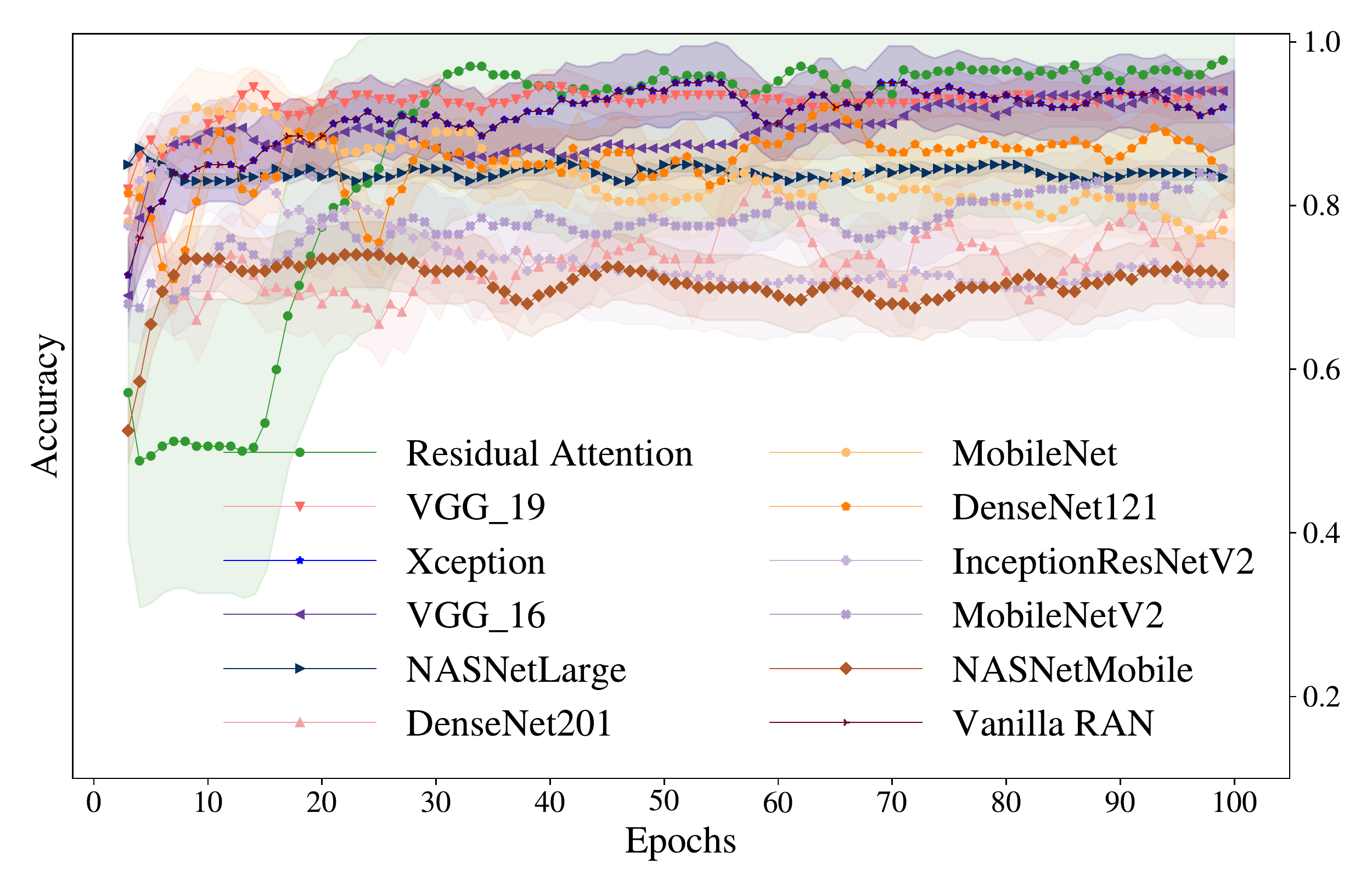}
		Accuracy over epochs of models on testing set
	\end{minipage}%
	\caption{Training and testing accuracy for all models. The shaded region represents 95\% confidence interval.}
	\label{fig:performance}
	\vspace{15pt}
\end{figure*}

\begin{table*}[ht]
	\centering
	\begin{tabular}[t]{>{\normalsize}l>{\normalsize}r>{\normalsize}r>{\normalsize}r>{\normalsize}r>{\normalsize}r>{\normalsize}r>{\normalsize}r>{\normalsize}r>{\normalsize}r>{\normalsize}r>{\normalsize}r>{\normalsize}r}

		\textbf{}&\multicolumn{7}{c}{\textbf{Testing Set}} & \multicolumn{5}{c}{\textbf{Validation Set}} \\
		\cmidrule{2-13}
		Network & Layers & Sens & Spec & Prec & Rec & Acc  & Sens & Spec & Prec & Rec  & Acc & AUC \\
		\cmidrule{1-13}
		NASNetMobile       	 & 198 & 0.72 & 0.68 & 0.69 & 0.70 & 0.7000  & 0.82 & 0.55 & 0.64 & 0.75 & 0.6818 & 0.86  \\		
		MobileNetV2	 		 & 54 & 0.72 & 1.00 & 1.00 & 0.78 & 0.8600  & 0.55 & 0.91 & 0.86 & 0.67 & 0.7273 & 0.72  \\
		InceptionResNetV2 	 & 246 & 0.40 & 1.00 & 1.00 & 0.63 & 0.7000 & 0.55 & 1.00 & 1.00 & 0.69 & 0.7727  & 0.72 \\
		DenseNet201          & 202 & 0.52 & 1.00 & 1.00 & 0.68 & 0.7600 & 0.82 & 0.91 & 0.90 & 0.83 & 0.8182  & 0.82 \\
		DenseNet121          & 122 & 0.92 & 0.72 & 0.77 & 0.90 & 0.8200 & 0.82 & 0.82 & 0.82 & 0.82 & 0.8636  & 0.90 \\	
		Xception             & 42 & 0.84 & 1.00 & 1.00 & 0.86 & 0.9200 & 0.91 & 0.82 & 0.83 & 0.90 & 0.8636  & 0.88 \\			
		NASNetLarge          & 270 & 0.72 & 0.92 & 0.90 & 0.77 & 0.8200 & 0.82 & 0.91 & 0.90 & 0.83 & 0.8636  & 0.90 \\		
		MobileNet            & 29 & 0.56 & 1.00 & 1.00 & 0.69 & 0.7800 & 0.82 & 1.00 & 1.00 & 0.85 & 0.9091  & 0.94 \\
		Vanilla Residual Att Net & 145 & 0.88 & 0.92 & 0.92 & 0.88 & 0.9000 & 0.91 & 0.91 & 0.91 & 0.91 & 0.9091  & 0.94 \\	
		VGG16                & 15 & 1.00 & 0.88 & 0.89 & 1.00 & 0.9400 & 1.00 & 0.82 & 0.85 & 1.00 & 0.9091  & 0.96 \\			
		VGG19                & 18 & 0.96 & 0.92 & 0.92 & 0.96 & 0.9400 & 0.92 & 1.00 & 1.00 & 0.92 & 0.9545 & 0.98 \\		
		\textbf{Residual Att Net (Our)} & 115 & 1.00 & 0.96 & 0.96 & 1.00 & \textbf{0.9800} & 1.00 & 1.00 & 1.00 & 1.00 & \textbf{1.0000} & 1.00  \\
		
		\hline
	\end{tabular}
	\vspace{15pt}
	\caption{Sensitivity, Specificity and Accuracy from all models on Testing and Validation set (sorted based on validation accuracy)}
	\label{tab:model_performance}
\end{table*}

\subsubsection{Models configurations}

Our selection of model size ranges from $15$ layers to $270$ layers deep. We trained each model with an initial learning rate of $1\mathrm{e}{-4}$ and a mini-batch size of 8 with 100 epochs. We used Adam as our optimizer. Adam optimizer uses two popular optimization techniques, Root Mean Square Propagation (RMSprop) and Stochastic Gradient Descent (SGD), with momentum. To provide stability during the training of our models, we used a learning rate decay with the decay rate shown in Equation~\ref{eq:decay_rate}. 
During the training, we used Binary Cross entropy as the loss function for all of the models. For all benchmark models, we use them with pre-trained weights using \textit{imagenet}, which contains $14$ million images with over $1000$ classes. Training on a large dataset requires massive computational power, for example, NASNetLarge was trained using $500$ GPUS for four days on the \textit{imagenet} dataset. We used the weights of models after the training and retrained them on our dataset. This methodology is commonly known as \textit{transfer learning}. 
\begin{equation}
\mbox{Decay Rate} = \frac{\mbox{Initial Learning Rate}}{\mbox{Epochs}}
\label{eq:decay_rate}
\end{equation}


\subsubsection{Output head modification}
The predefined models have been designed and trained for the very large \textit{imagenet} dataset. This dataset consists of of $1000$ classes and millions of images whereas our dataset has two classes and $239$ images. To prevent overfitting of the predefined models on our small dataset, we modified the output layers of all of the models. We removed the output layer and added a custom output head. For example, VGG16 has three fully connected layers as the output head, the first two layers have $4096$ neurons while the third has $1000$ (number of classes) neurons.  We modified this output from three layers to two layers with $64$ neurons in the first layer and $2$ (number of classes) neurons in the second layer. 
We did this on all of the models to standardize the output head.

\begin{figure*}[ht]
	\centering
	\begin{subfigure}{0.5\textwidth}
		\centering
		\includegraphics[width=.9\linewidth]{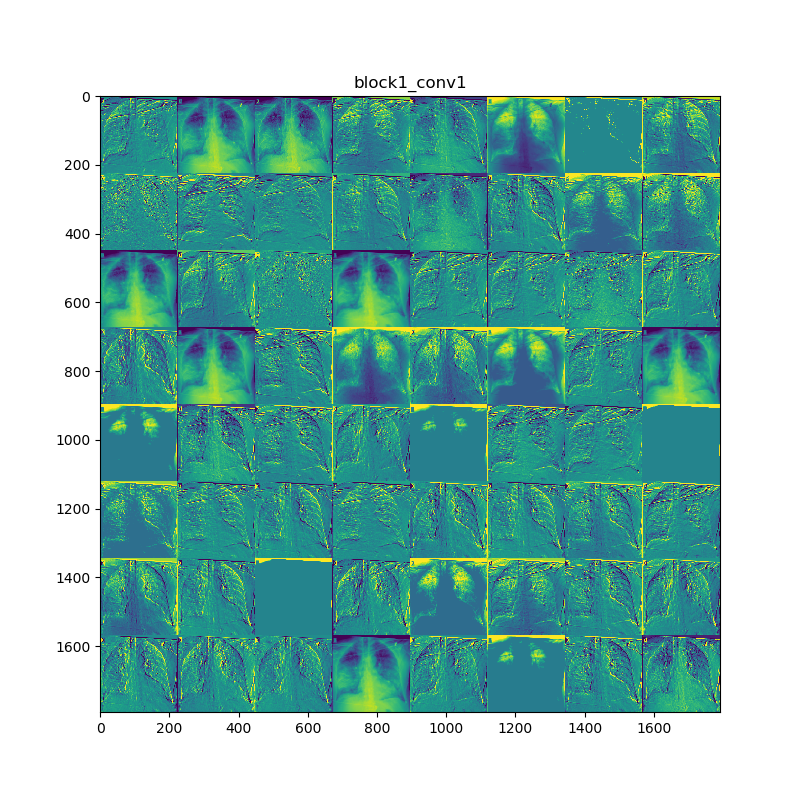}
		\vspace{-10pt}
		\caption{Feature Map at Convolution Layer 1}
		\label{fig:feature_map_1}
	\end{subfigure}%
	\hspace{-40pt}
	\begin{subfigure}{0.5\textwidth}
		\centering
		\includegraphics[width=0.9\linewidth]{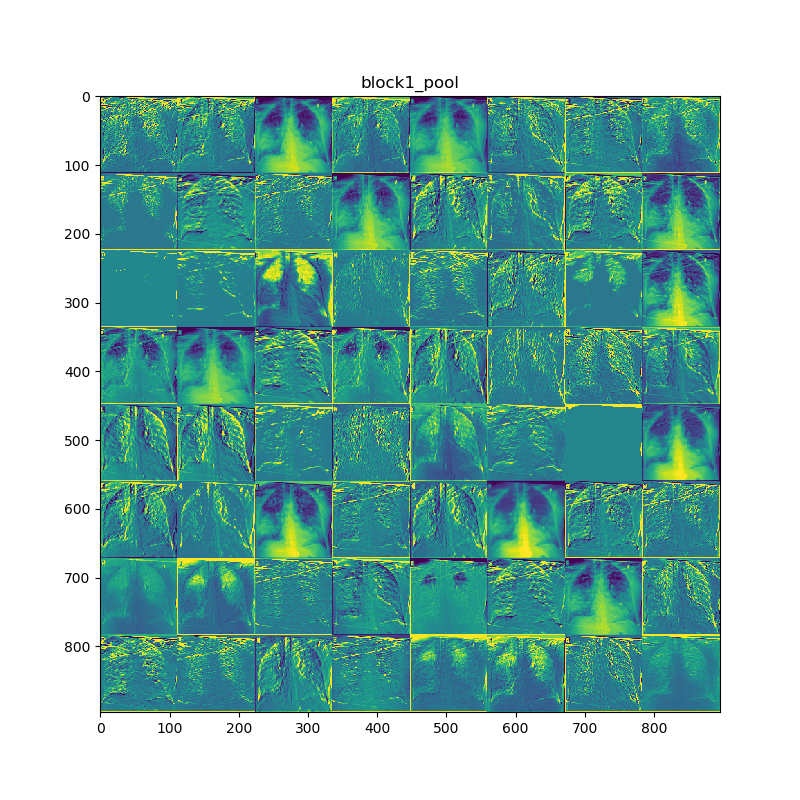}
		\vspace{-10pt}
		\caption{Feature Map at pooling Layer}
		\label{fig:feature_map_2}
	\end{subfigure}
	\vspace{10pt}
	\caption{Feature maps of Residual Attention Network at first 2 Layers}
	\label{fig:feature_map}
\end{figure*}

\subsection{Results}
Table~\ref{tab:model_performance} displays the accuracy of the models on the testing and validation sets of our COVID-19 dataset. We observe that our custom Residual Attention Network outperforms all other deep learning models, while NASNetMobile turns out to be the worst performer. Our experiments show that VGG19 performs better than VGG16. Even though both models are similar, VGG19 has more layers and it is commonly thought that deeper pre-trained models perform better than shallow models. But that is not the case for our dataset, for instance, the network with a shallower depth, DenseNet121, performed better than a very deep network, DenseNet201 (both networks are at least six times deeper than VGG19). From this, we observe that it boils down to feature maps of layers from a model, which is dependent on the convolution layers. We also observe that very deep networks do not perform better than shallower ones, since the top two performing networks are shallow networks ($1/4$ of the size of the largest network). Figure~\ref{fig:performance} shows the model training and testing accuracy for every epoch. We observe large/deeper models (DenseNet201, NASNetLarge, InceptionResNetV2) tend to overfit and shallower models (VGG16, VGG19, Residual Attention Network) are able to generalize our dataset better. We also observe our residual attention model during testing took ~$15$ epochs to start improving accuracy. This can be explained by our layers weight initializer using Xavier where it initializes the weights of the layer with zero mean and unit variance.

Overall, our experiments show that the Residual Attention Network performs best: with $98$\% accuracy on testing and $100$\% on validation to screen COVID-19 in X-ray images. 

\textit{COVID-19 prediction explanation: }
Figure~\ref{fig:feature_map} visualizes the feature maps of the first two layers of our Residual Attention Network model trained on our COVID-19 dataset. There are $64$ images in each picture with an $8\times8$ grid for each image. The feature maps were generated by passing an image to our trained model and collecting data of activated neurons. 
We select only the first two layers and $64$ neurons of each layer, since the number of neurons and parameters grows exponentially at later layers. We observe from the feature maps of both layers that our model can extract relevant details from an image, \eg lungs, spine, veins, potential COVID-19 affected areas.  

\section{Conclusion}
\label{sec:conclusion}

This paper proposes a novel method to detect COVID-19 using chest X-ray images.  The method uses a Residual Attention Network and data augmentation. We collected and curated a dataset of $239$ images: $120$ images of patients infected with COVID-19, and $119$ images of non-COVID-19 patients, which we label ``normal.'' The dataset is diverse in terms of patient age, gender, and location. We applied a non-linear dimensionality reduction technique, UMAP, and observed a clear distinction between X-rays of COVID-19 and normal patients. We designed and implemented a Residual Attention Network to classify COVID-19 patients and compared our model with many popular deep learning models: VGG, DenseNet, NASNet, Xception, and Inception. Our experiments show that our custom Residual Attention Network performs best among all of the models with $98$\% testing and $100$\% validation accuracy. We generated feature maps of the Residual Attention Network and they show that the low-level features extracted from a given image include areas of potential COVID-19 infection.

This research shows that chest X-ray images can potentially be used to detect COVID-19, or can be combined with other testing methods to corroborate a diagnostic outcome. 
The immediate future work is to add more images to our dataset. 
The second avenue of future work is to work with domain experts to study the utility of deploying the technique in practice. Due to the recent nature of the COVID-19 pandemic, the deployment has lagged behind the research.
Another aspect of future work is to study whether chest X-rays can be used to detect COVID-19 early in the illness or in asymptomatic cases.
Our X-ray dataset does not have information about patient symptoms or time with the disease, only that the X-rays are of COVID-19 positive patients.
Since early detection is important to effective quarantining, it would be interesting to test our model on X-rays from patients who have just been infected or who are asymptomatic.




\section*{Acknowledgments}
This work was supported in part by the National Science Foundation under Award No.~1759965, \textit{Collaborative Research: ABI Development: Symbiota2: Enabling greater collaboration and flexibility for mobilizing biodiversity data}. Opinions, findings and conclusions or recommendations expressed in this material are those of the author(s) and do not necessarily reflect those of the National Science Foundation.

\bibliographystyle{ACM-Reference-Format}
\bibliography{bib}



\end{document}